\documentclass[12pt]{article}
\usepackage{amsmath}
\usepackage{amssymb}

\begin{document}

\renewcommand{\theequation}
{\thesection.\arabic{equation}}
\thispagestyle{empty}
\vspace*{20mm} 
\begin{center}
{\LARGE {\bf UV/IR Mixing and Anomalies }} \\
 \qquad \\
{\LARGE {\bf in Noncommutative Gauge Theories}} \\

\vspace*{20mm}

\renewcommand{\thefootnote}{\fnsymbol{footnote}}

{\Large Tadahito NAKAJIMA 
\footnote[1]{E-mail address: nakajima@phys.ge.cst.nihon-u.ac.jp}} \\
\vspace*{20mm}

{\large {\it Laboratory of Physics, College of Science and Technology, 
Nihon University}} \\
{\large {\it Narashino-dai, Funabashi, Chiba 274-8501, Japan}} \\ 

\vspace*{20mm}

{\bf Abstract} \\

\end{center}

Using path integral method (Fujikawa's method) we calculate anomalies in 
noncommutative gauge theories with fermions in the bi-fundamental and adjoint 
representations. We find that axial and chiral gauge anomalies coming from 
non-planar contributions are derived in the low noncommutative momentum 
limit $\widetilde{p}^{\mu}(\equiv \theta^{\mu\nu}p_{\nu}) \rightarrow 0$. The 
adjoint chiral fermion carries no anomaly in the non-planar sector in $D=4k
(k=1,2,\ldots,)$ dimensions. It is naturally shown from the path integral 
method that anomalies in non-planar sector originate in UV/IR mixing. 

\vspace*{15mm}

\clearpage
\setcounter{section}{0}
\section{Introduction}
\setcounter{page}{1}
\setcounter{equation}{0}

Gauge theories on noncommutative space (or simply noncommutative gauge 
theories) are the subject of much recent interest. (See for recent reviews 
\cite{MRDNAN, RJS} and references therein.) One of the interesting 
feature of noncommutative gauge theories at the quantum level is UV/IR 
mixing \cite{SMMVR}. 
Although the planar diagrams are essentially the same as those in the 
corresponding ordinary field theories, the non-planar diagrams can also be 
seen to exhibit an interesting stringy phenomenon. The planar diagrams 
control the UV properties, while the non-planar diagrams generally lead to 
new IR phenomena through the mixing.

Anomalies in noncommutative gauge theories have been discussed by several 
authors \cite{FANS0}-\cite{AAELST}. There are two kinds of anomalies one of 
which comes from the planar contributions and the other of which from 
the non-planar contributions. The anomalies in planar 
sector can be evaluated by several different methods. Axial anomalies have 
been calculated by the path integral formulation (Fujikawa's method) besides 
the perturbative analysis \cite{FANS0, JMGBCPM}. Chiral gauge anomalies 
can also be described using cohomological methods \cite{LBMSAT, LBAS}. 
It is known that these anomalies take the form of the straightforward Moyal 
deformation in the corresponding anomalies in ordinary gauge theories. The 
theta (noncommutative) parameter does not explicitly appear in the final 
formula except in the appearance of the Moyal star product.

The anomalies in non-planar sector have been studied from different points 
of view \cite{CPM, FANS, KIJK, AAELST}. When the chiral 
fermions of the noncommutative gauge theories are in the bi-fundamental 
representation and the adjoint representation, there are non-planar 
contributions from the non-planar diagrams. For the non-vanishing 
noncommutative momentum $\widetilde{p}^{\mu} 
\equiv \theta^{\mu\nu}p_{\nu}$, however, the non-planar triangle diagrams can 
be expressed in terms of the (modified) Bessel functions and they are 
UV-finite without any regularizations. Hence, there are no anomalous 
contributions from the non-planar diagrams \cite{CPM, KIJK}. 
On the other hand, it was shown that axial anomaly in non-planar sector does 
not vanish. For noncommutative QED with fermions in the fundamental 
representation, there are two kinds of axial currents in which the order of 
the product of the fermions differs. One of these currents leads to the 
anomaly of the non-planar contributions when the noncommutative momentum 
is very small \cite{FANS}. Anomalies in non-planar sector in the case of zero 
noncommutative momentum have been discussed in detail in \cite{AAELST}.

These arguments on the anomalies in non-planar sector are based 
on the perturbative analysis, while the anomalies in planar sector can be 
evaluated by the path integral method and cohomological approach besides the 
perturbative analysis. Therefore, it will be natural to consider  
approaches other than the perturbative analyze in the evaluation of the 
anomalies in non-planar sector. In this paper we would like to derive axial 
and chiral gauge anomalies in non-planar sector by path integral method. 
The path integral method will be found to be suited for the calculation of 
the anomalies in non-planar sector. The paper is organized as follows. 
In Sec. 2, we consider a noncommutative gauge theory with fermion in 
the bi-fundamental and the adjoint representation and derive the axial 
anomaly in non-planar sector by the path integral method. In Sec. 3, the 
path integral method is also applied in deriving the chiral gauge anomaly in 
a noncommutative chiral gauge theory with chiral fermion in the bi-fundamental 
and the adjoint representation. Sec. 4 is devoted to conclusions.

%
%
\setcounter{section}{1}
\section{The axial anomaly for bi-fundamental and adjoint fermion}
\setcounter{equation}{0}
We first discuss the gauge theories with fermions in the bi-fundamental 
representation in noncommutative Euclidean space. 
Let us consider a bi-fundamental Dirac fermion $\psi^{i}{}_{j}(x)$ interacting 
with a $U(N_{A})$ gauge field $A_{\mu}{}^{i_{1}}{}_{i_{2}}(x)$ and a 
$U(N_{B})$ gauge field $B_{\mu}{}^{j_{2}}{}_{j_{1}}(x)$. 
Here the index $i$ runs from 1 to $N_{A}$ and $j$ from 1 to $N_{B}$, 
respectively. The classical action of this theory on 2n-dimensional 
noncommutative (Euclidean) space is given by
\begin{eqnarray}
S[\bar{\psi}, \psi, A, B] = \int d^{2n}x \; \bar{\psi}^{j}{}_{i}(x) \ast 
(i /\!\!\!\!D[A, B]) \psi^{i}{}_{j} (x) \;.
\end{eqnarray}
%
Here the operator $/\!\!\!\!D[A, B]$ denotes the Dirac operator whose 
concrete form is given by, 
\begin{eqnarray}
& & /\!\!\!\!D[A, B]\psi^{i_{1}}{}_{j_{1}}(x) = 
/\!\!\!\partial \psi^{i_{1}}{}_{j_{1}}(x) 
+ A_{\mu}{}^{i_{1}}{}_{i_{2}}(x) \ast \gamma^{\mu} \psi^{i_{2}}{}_{j_{1}}(x) 
- \gamma^{\mu} \psi^{i_{1}}{}_{j_{2}}(x) \ast B_{\mu}{}^{j_{2}}{}_{j_{1}}(x) 
\nonumber \\ 
& & = (\; /\!\!\!\partial \delta^{i_{1}}{}_{i_{2}}\delta^{j_{1}}{}_{j_{2}} 
+ A_{\mu}{}^{i_{1}}{}_{i_{2}}(x) \ast \delta^{j_{1}}{}_{j_{2}} \gamma^{\mu}
- \delta^{i_{1}}{}_{i_{2}} \ast B_{\mu}{}^{j_{1}}{}_{j_{2}}(x) \gamma^{\mu} 
\;) \psi^{i_{2}}{}_{j_{2}}(x) \;, 
\end{eqnarray}
%
with the notations $(A_{\mu} \ast ) \psi \equiv A_{\mu} \ast \psi$ and 
$(\ast B_{\mu} ) \psi \equiv \psi \ast B_{\mu}$ \cite{CPM2}. Since we have 
chosen the gamma matrices $\gamma^{\mu}$, $\mu =1,2,\ldots, 2n$ as 
Hermitian matrices, the matrix $\gamma_{2n+1} \equiv (-i)^{n}\prod^{2n}_{k=1} 
\gamma_{\mu_{k}}$ remains Hermitian. The symbol $\ast$ stands for the Moyal 
star product defined as follows, 
\begin{eqnarray}
f(x) \ast g(x) \!\!&=&\!\! \left.
e^{\frac{i}{2}\theta^{\mu\nu}
\frac{\partial}{\partial \xi_{\mu}}\frac{\partial}{\partial \zeta_{\nu}}}
f(x+\xi)g(x+\zeta) \right|_{\xi=\zeta=0} \nonumber \\
\!\!&=&\!\! \int \frac{d^{4}p}{(2\pi)^{4}}\int \frac{d^{4}q}{(2\pi)^{4}}
e^{-\frac{i}{2}p_{\mu}\theta^{\mu\nu}q_{\nu}}
e^{i(p_{\mu}+q_{\mu})x^{\mu}} \widetilde{f}(p)\widetilde{g}(q) \;,
\end{eqnarray}
%
where $\theta^{\mu\nu}=-\theta^{\nu\mu}$ denotes an antisymmetric real matrix. 

We begin with the evaluation of the axial anomalies. 
We perform a infinitesimal (local) chiral transformation: 
\begin{eqnarray}
& & \delta_{2n+1} \psi(x){}^{i}{}_{j} = 
i\lambda_{A}(x) \ast 
\gamma_{2n+1} \psi(x){}^{i}{}_{j} 
- i\gamma_{2n+1} \psi(x){}^{i}{}_{j} \ast 
\lambda_{B}(x) \;, \nonumber \\ 
& & \\
& & \delta_{2n+1} \bar{\psi}{}^{j}{}_{i}(x) = 
i\bar{\psi}{}^{j}{}_{i}(x) \gamma_{2n+1} \ast \lambda_{A}(x)
- i\lambda_{B}(x) \ast 
 \bar{\psi}{}^{j}{}_{i}(x) \gamma_{2n+1} \;,  \nonumber 
\end{eqnarray}
%
where $\lambda_{A}(x)$ and $\lambda_{B}(x)$ denote some infinitesimal 
functions. For the infinitesimal transformation, the action (2.1) changes to 
\begin{eqnarray}
& & \delta_{2n+1} S = 
- \int d^{2n}x \left\{ \lambda_{A} \ast D^{(A)}_{\mu}j^{\mu (A)}_{2n+1} 
+ \lambda_{B} \ast D^{(B)}_{\mu}j^{\mu (B)}_{2n+1} \right\} \;,  \nonumber 
\end{eqnarray}
where the currents $j^{\mu (A)}_{2n+1}(= j^{\mu (A)}_{2n+1}{}^{i}{}_{i})$ and 
$j^{\mu (B)}_{2n+1}(= j^{\mu (B)}_{2n+1}{}^{j}{}_{j})$ are defined, 
respectively, by the 
identities 
\begin{eqnarray}
& & j^{\mu (A)}_{2n+1}(x) \equiv 
(\psi_{\beta}{}^{i}{}_{j}(x) \ast \bar{\psi}_{\alpha}{}^{j}{}_{i}(x))
(\gamma^{\mu}\gamma_{2n+1})_{\alpha\beta} \;,  \nonumber \\ 
& & \\
& & j^{\mu (B)}_{2n+1}(x) \equiv 
(\bar{\psi}_{\alpha}{}^{j}{}_{i}(x) \ast \psi_{\beta}{}^{i}{}_{j}(x))
(\gamma^{\mu}\gamma_{2n+1})_{\alpha\beta} \;,  \nonumber 
\end{eqnarray}
%
and the covariant derivative of these current are given by 
\begin{eqnarray}
& & D^{(A)}_{\mu}j^{\mu (A)}_{2n+1}(x) \equiv 
\partial_{\mu} j^{\mu (A)}_{2n+1}(x) + A_{\mu}{}^{i_{1}}{}_{i_{2}}(x) 
\ast j^{\mu (A)}_{2n+1}{}^{i_{2}}{}_{i_{1}}(x) 
- j^{\mu (A)}_{2n+1}{}^{i_{1}}{}_{i_{2}}(x) \ast 
A_{\mu}{}^{i_{2}}{}_{i_{1}}(x)  \;, \nonumber \\ 
& & \nonumber \\
& & D^{(B)}_{\mu}j^{\mu (B)}_{2n+1}(x) \equiv 
\partial_{\mu} j^{\mu (B)}_{2n+1}(x) + B_{\mu}{}^{j_{1}}{}_{j_{2}}(x) 
\ast j^{\mu (B)}_{2n+1}{}^{j_{2}}{}_{j_{1}}(x) 
- j^{\mu (B)}_{2n+1}{}^{j_{1}}{}_{j_{2}}(x) \ast 
B_{\mu}{}^{j_{2}}{}_{j_{1}}(x)  \;. \nonumber  
\end{eqnarray}
The partition function with classical action given in the expression (2.1) is 
defined as 
\begin{eqnarray} 
Z [A, B] = \int {\cal D} \bar{\psi} {\cal D} \psi \exp \left( 
-S[\bar{\psi}, \psi, A, B] \right)\;.  \nonumber 
\end{eqnarray}
In order to compute the change of the path integral measure under the chiral 
transformation (2.4), we introduce an orthonormal and complete set of 
eigenfunctions $\{ \varphi_{n} \}$ of the Dirac operator $/\!\!\!\!D[A, B]$. 
%
%
%
The fermions can be expanded in the orthonormal basis of 
eigenfunctions $\{ \varphi_{n} \}$ as $\psi(x) = \sum_{n}a_{n}\varphi_{n}(x)$ 
and $\bar{\psi}(x) = \sum_{n}\bar{b}_{n} 
\varphi_{n}^{\dagger}(x)$, where the coefficients $a_{n}$ and $\bar{b}_{n}$ 
are Grassmann numbers. 
Under the infinitesimal transformations (2.4), the integration measure of the 
fermionic fields transform as 
${\cal D}\widetilde{\psi}{\cal D}\widetilde{\bar{\psi}} = 
J_{axial}[\lambda_{A}, \lambda_{B}]{\cal D}\psi {\cal D}\bar{\psi}$ with the 
Jacobian, 
%
%
%
%
\begin{eqnarray} 
J_{axial}[\lambda_{A}, \lambda_{B}] 
= \exp\left( -2i {\cal A}_{axial}[\lambda_{A}, \lambda_{B}] 
\right) \;, 
\end{eqnarray}
%
where ${\cal A}_{axial}$ includes the sum over the eigenstates $n$. 
Since the sum is ill-defined, we must regularize the sum in a gauge invariant 
way. This is done by introducing a 
Gaussian damping factor, 
%
%
\begin{eqnarray} 
{\cal A}^{\rm reg}_{axial}[A, B, \lambda_{A}, \lambda_{B}] 
\!\!&=&\!\! \lim_{\varepsilon \longrightarrow 0} 
\Sigma_{n} \int d^{2n}x \Bigl[ \Bigr. \,
\lambda_{A}(x) \ast \gamma_{2n+1} e^{-\varepsilon /\!\!\!\!D[A, B]^{2}} 
\varphi_{n}(x) \ast \varphi^{\dagger}_{n}(x) \nonumber \\
\!\!&-&\!\! \lambda_{B}(x) \ast \varphi^{\dagger}_{n}(x) \ast \gamma_{2n+1}  
e^{-\varepsilon /\!\!\!\!D[A, B]^{2}} \varphi_{n}(x) 
\, \Bigl. \Bigr] \;.
\end{eqnarray}
%
%
%
%
%
%
%
We now evaluate the regularized sum in Fourier space 
$\displaystyle{\varphi_{n}(x)=\int\frac{d^{2n}k}{(2\pi)^{2n}}e^{ik\cdot x} 
\widetilde{\varphi}_{n}(k)}$. Then we have 
\begin{eqnarray} 
& & {\cal A}^{\rm reg}_{axial}[A, B, \lambda_{A}, \lambda_{B}] \nonumber \\
& & = \lim_{\varepsilon \longrightarrow 0} 
\int d^{2n}x \int \frac{d^{2n}k}{(2\pi)^{n}} 
\mbox{\boldmath ${\rm Tr}$} 
\Bigl[ \Bigr. \, \lambda_{A}(x) \ast \gamma_{2n+1} 
\exp\left(-\varepsilon \left\{ (i/\!\!\!k+/\!\!\!\!D[A, B])^{2}
\right\} \right) e^{ik \cdot x} \ast e^{-ik \cdot x} \nonumber \\
& & - \lambda_{B}(x) \ast  e^{-ik \cdot x} \ast \gamma_{2n+1}  
\exp\left(-\varepsilon \left\{ (i/\!\!\!k+/\!\!\!\!D[A, B])^{2}
\right\} \right) e^{ik \cdot x} \,
\Bigl. \Bigr] \,  \;,
\end{eqnarray}
%
where $k \cdot x \equiv k_{\mu}x^{\mu}$ and the notation 
``$\mbox{\boldmath ${\rm Tr}$}$" denotes 
the trace over the $U(N_{A})$, $U(N_{B})$ and the Dirac matrices. Notice that 
multiplication by a plane wave $e^{ik \cdot x}$ translates a 
general function as $e^{-ik \cdot x} \ast f(x) \ast e^{ik \cdot x}=
f(x-\widetilde{k})$, where $\widetilde{k}^{\mu} \equiv 
\theta^{\mu\nu}k_{\nu}$. This exhibits that 
large momenta will lead to large nonlocality of the theory. Taking this into 
account and inserting decomposition $/\!\!\!\!D^{2} = D_{\mu}D^{\mu} 
+ \frac{1}{2}\gamma^{\mu}\gamma^{\nu} [\,D_{\mu}, \, D_{\nu} \,]$ into the 
expression (2.8), we obtain 
\begin{eqnarray} 
& & {\cal A}^{\rm reg}_{axial}[A, B, \lambda_{A}, \lambda_{B}] = 
\lim_{\varepsilon \longrightarrow 0} 
\int d^{2n}x \int \frac{d^{2n}k}{(2\pi)^{n}} e^{\varepsilon k \cdot k} \\
& &  \mbox{\boldmath ${\rm Tr}$} 
\Bigl[ \Bigr. \, \lambda_{A}(x) \ast \gamma_{2n+1} 
\exp\Bigl( \Bigr. \,-\varepsilon \Bigl\{ \Bigr.
2ik_{\mu} \left( \partial^{\mu} + A^{\mu}(x) \ast - 
 \ast B^{\mu}(x+\widetilde{k}) \right) \nonumber \\
& & + \left( \partial^{\mu} + A^{\mu}(x) \ast - 
 \ast B^{\mu}(x+\widetilde{k}) \right)^{2} 
 +\frac{1}{2}\gamma^{\mu}\gamma^{\nu} 
 \left( F^{A}_{\mu\nu}(x) \ast - \ast F^{B}_{\mu\nu}(x+\widetilde{k}) \right) 
\Bigl. \Bigr\} \Bigl. \Bigr) \nonumber \\
& & - \lambda_{B}(x) \ast \gamma_{2n+1} 
\exp\Bigl( \Bigr. \,-\varepsilon \Bigl\{ \Bigr.
2ik_{\mu} \left( \partial^{\mu} + A^{\mu}(x-\widetilde{k}) \ast - 
 \ast B^{\mu}(x) \right) \nonumber \\
& & + \left( \partial^{\mu} + A^{\mu}(x-\widetilde{k}) \ast - 
 \ast B^{\mu}(x) \right)^{2} 
 +\frac{1}{2}\gamma^{\mu}\gamma^{\nu}
 \left( F^{A}_{\mu\nu}(x-\widetilde{k}) \ast - \ast F^{B}_{\mu\nu}(x) \right) 
\Bigl. \Bigr\} \Bigl. \Bigr) \, \Bigl. \Bigr]  \;, \nonumber
\end{eqnarray}
%
with $F^{A}_{\mu\nu} \equiv \partial_{\mu}A_{\nu}-\partial_{\nu}A_{\mu}
+[\,A_{\mu}, \,A_{\nu}\,]_{\ast}$ and $F^{B}_{\mu\nu} \equiv 
\partial_{\mu}B_{\nu}-\partial_{\nu}B_{\mu}+[\,B_{\mu}, \,B_{\nu}\,]_{\ast}$. 
%
%
%
%
We expand the exponential and utilize the trace properties of the Dirac 
matrices: \\ ${\rm Tr}(\gamma^{2n+1}\gamma^{\mu_{1}}\gamma^{\mu_{2}} 
\cdots \gamma^{\mu_{2n}}) = (-2i)^{n}\varepsilon^{\mu_{1}\mu_{2} 
\cdots \mu_{2n}}$, where $\varepsilon^{\mu_{1}\mu_{2} \cdots \mu_{2n}}$ is 
the Levi-Civita tensor. 
Then only the term of n-th order in $F_{\mu\nu}$ 
remains under the limit $\varepsilon \longrightarrow 0$. Performing the 
rescaling $k_{\mu} \longrightarrow (1/\sqrt{\varepsilon})k_{\mu}$, we have  
\begin{eqnarray} 
& & {\cal A}^{\rm reg}_{axial}[A, B, \lambda_{A}, \lambda_{B}] = 
\lim_{\varepsilon \longrightarrow 0} 
\int d^{2n}x \int \frac{d^{2n}k}{(2\pi)^{n}} e^{k \cdot k}
\frac{i^{n}}{n!} \varepsilon^{\mu_{1}\mu_{2} \cdots \mu_{2n-1}\mu_{2n}} \\
& &  \Bigl\{ \Bigr.
\lambda_{A}(x) \ast {\rm Tr}_{A}{\rm Tr}_{B} \left[ \, 
(F^{A}_{\mu_{1}\mu_{2}}(x) \ast 
- \ast F^{B}_{\mu_{1}\mu_{2}}(x+\frac{\widetilde{k}}{\sqrt{\varepsilon}}))
 \cdots (F^{A}_{\mu_{2n-1}\mu_{2n}}(x) \ast 
-  \ast F^{B}_{\mu_{2n-1}\mu_{2n}}(x+\frac{\widetilde{k}}{\sqrt{\varepsilon}})) 
\right] \nonumber \\ 
& & - \lambda_{B}(x) \ast {\rm Tr}_{A}{\rm Tr}_{B} \left[ \, 
(F^{A}_{\mu_{1}\mu_{2}}(x-\frac{\widetilde{k}}{\sqrt{\varepsilon}}) \ast 
- \ast F^{B}_{\mu_{1}\mu_{2}}(x)) \cdots 
(F^{A}_{\mu_{2n-1}\mu_{2n}}(x-\frac{\widetilde{k}}{\sqrt{\varepsilon}}) 
\ast - \ast F^{B}_{\mu_{2n-1}\mu_{2n}}(x)) 
\right]   \Bigl. \Bigr\} \;, \nonumber 
\end{eqnarray}
%
where the notations ${\rm Tr}_{A}$ and ${\rm Tr}_{B}$ denote the traces over 
the $U(N_{A})$ and $U(N_{B})$ matrices, respectively. 

Before advancing the calculation of axial anomaly in arbitrary dimensions, 
we will examine the case of two and four dimensions concretely. 

%
%
\begin{flushleft}
{\bf  Two dimensions}
\end{flushleft}

First we consider the case of two dimensions. The explicit form of the 
expression (2.10) is given by 
\begin{eqnarray} 
{\cal A}^{\rm reg}_{axial}[A, B, \lambda_{A}, \lambda_{B}] \!\!&=&\!\! 
\lim_{\varepsilon \longrightarrow 0} 
\int d^{2}x \int \frac{d^{2}k}{(2\pi)^{2}} e^{k \cdot k}
i \varepsilon^{\mu\nu} 
\Bigl\{ \Bigr. \,
N_{B}\, \lambda_{A}(x) \ast {\rm Tr}_{A} F^{A}_{\mu\nu}(x) 
+ N_{A} \,\lambda_{B}(x) \ast {\rm Tr}_{B} F^{B}_{\mu\nu}(x) \nonumber \\ 
\!\!&-&\!\! N_{B}\, \lambda_{A}(x) \ast 
{\rm Tr}_{A}(F^{A}_{\mu\nu}(x-\frac{\widetilde{k}}{\sqrt{\varepsilon}})
-N_{A} \, \lambda_{B}(x) \ast 
{\rm Tr}_{B}F^{B}_{\mu\nu}(x+\frac{\widetilde{k}}{\sqrt{\varepsilon}}) 
\Bigl. \Bigr\} \;,
\end{eqnarray}
%
where the coefficients $N_{A} (\equiv {\rm Tr}_{A} I_{N_{A} \times N_{A}})$ 
and $N_{B} (\equiv {\rm Tr}_{B} I_{N_{B} \times N_{B}})$ come from the trace 
of the unit matrices $I_{N_{A} \times N_{A}}$ and $I_{N_{B} \times N_{B}}$, 
respectively. Performing integration over the momentum $k_{\mu}$, we find 
\begin{eqnarray} 
{\cal A}^{\rm reg}_{axial}[A, B, \lambda_{A}, \lambda_{B}] \!\!&=&\!\! 
\lim_{\varepsilon \longrightarrow 0} 
\int d^{2}x \frac{1}{4\pi} 
i \varepsilon^{\mu\nu} 
\Bigl\{ \Bigr. \,
N_{B}\, \lambda_{A}(x) \ast {\rm Tr}_{A} F^{A}_{\mu\nu}(x) 
+ N_{A} \,\lambda_{B}(x) \ast {\rm Tr}_{B} F^{B}_{\mu\nu}(x) \nonumber \\ 
\!\!&-&\!\! N_{B}\, \lambda_{A}(x) \ast 
 {\rm Tr}_{A} \int \frac{d^{2}p}{(2\pi)^{2}} 
 \widetilde{F}^{A}_{\mu\nu}(p)e^{ip \cdot x}
 \exp\left(\frac{1}{4\varepsilon} p \circ p \right) \nonumber \\ 
\!\!&-&\!\! N_{A} \, \lambda_{B}(x) \ast 
{\rm Tr}_{B} \int \frac{d^{2}p}{(2\pi)^{2}} 
\widetilde{F}^{B}_{\mu\nu}(p)e^{ip \cdot x}
 \exp\left(\frac{1}{4\varepsilon} p \circ p \right)
\Bigl. \Bigr\} \;,
\end{eqnarray}
%
where $\widetilde{F}_{\mu\nu}(p)$ is the Fourier coefficients of the field 
strength ${F}_{\mu\nu}(x)$ and the notation $p \circ p \equiv 
\eta_{\rho\sigma}p_{\mu}\theta^{\mu\rho}p_{\nu}\theta^{\nu\sigma}
(=\widetilde{p}_{\mu}\widetilde{p}^{\mu})$ denotes 
the square of the noncommutative momentum \cite{SMMVR}. 
It can be regarded that the third and fourth term in the right-hand side 
of the expression (2.12) come from the non-planar contributions, while the 
first and the second term come from the planar contributions. The quantity 
$p \circ p$ satisfies the condition $p \circ p \leq 0$ under the Euclidean 
metric: ${\rm diag}\,\eta_{\mu\nu} = (-1, -1)$ and hence the factor 
$\exp\left(\frac{1}{4\varepsilon} p \circ p \right)$ plays the role 
of the damping factor. Since it satisfies $p \circ p < 0$ for arbitrary 
non-zero momentum $p_{\mu}$ in two dimensions, the expression (2.12) becomes 
as follows 
\begin{eqnarray} 
{\cal A}^{\rm reg}_{axial}[A, B, \lambda_{A}, \lambda_{B}] \!\!&=&\!\! 
\int d^{2}x \frac{1}{4\pi} 
i \varepsilon^{\mu\nu} 
\Bigl\{ \Bigr. \,
N_{B}\, \lambda_{A}(x) \ast {\rm Tr}_{A} F^{A}_{\mu\nu}(x) 
+ N_{A} \,\lambda_{B}(x) \ast {\rm Tr}_{B} F^{B}_{\mu\nu}(x) \nonumber \\ 
\!\!&-&\!\! 
N_{B}\, \lambda_{A}(x) {\rm Tr}_{A} \widetilde{F}^{A}_{\mu\nu}(0) 
-N_{A} \, \lambda_{B}(x) {\rm Tr}_{B} \widetilde{F}^{B}_{\mu\nu}(0) 
\Bigl. \Bigr\} \;,
\end{eqnarray}
%
by taking the limit $\varepsilon \longrightarrow 0$. If the Fourier 
coefficients $\widetilde{F}^{A}_{\mu\nu}(0)$ and 
$\widetilde{F}^{B}_{\mu\nu}(0)$ are non-zero, then there are the non-planar 
contributions to the axial anomaly \cite{AAELST}. 
Note that the terms coming from the non-planar sector in the expression (2.13) 
do not diverge under the ``local" chiral transformations.

%
\begin{flushleft}
{\bf Four dimensions}
\end{flushleft}

Next we consider the case of four dimensions. The explicit form of the 
expression (2.10) is given by 
\begin{eqnarray} 
& & {\cal A}^{\rm reg}_{axial}[A, B, \lambda_{A}, \lambda_{B}] = 
\lim_{\varepsilon \longrightarrow 0} 
\int d^{4}x \int \frac{d^{4}k}{(2\pi)^{4}} e^{k \cdot k}
\frac{-1}{2} \varepsilon^{\mu\nu\rho\sigma} \\
& &  \Bigl\{ \Bigr.
\lambda_{A}(x) \ast {\rm Tr}_{A}{\rm Tr}_{B} \left[ \, 
(F^{A}_{\mu\nu}(x) \ast - 
\ast F^{B}_{\mu\nu}(x+\frac{\widetilde{k}}{\sqrt{\varepsilon}}))
(F^{A}_{\rho\sigma}(x) \ast 
-  \ast F^{B}_{\rho\sigma}(x+\frac{\widetilde{k}}{\sqrt{\varepsilon}})) 
\right] \nonumber \\ 
& & - \lambda_{B}(x) \ast {\rm Tr}_{A}{\rm Tr}_{B} \left[ \, 
(F^{A}_{\mu\nu}(x-\frac{\widetilde{k}}{\sqrt{\varepsilon}}) \ast 
- \ast F^{B}_{\mu\nu}(x)) 
(F^{A}_{\rho\sigma}(x-\frac{\widetilde{k}}{\sqrt{\varepsilon}}) \ast 
-  \ast F^{B}_{\rho\sigma}(x)) 
\right]   \Bigl. \Bigr\} \;. \nonumber 
\end{eqnarray}
%
Performing the integration over the momentum $k_{\mu}$, we find 
\begin{eqnarray} 
& & {\cal A}^{\rm reg}_{axial}[A, B, \lambda_{A}, \lambda_{B}] = 
-\lim_{\varepsilon \longrightarrow 0} 
\int d^{4}x \frac{1}{32\pi^{2}} 
\varepsilon^{\mu\nu\rho\sigma}  \\
& & \Bigl\{ \Bigr. \,
N_{B}\, \lambda_{A}(x) \ast 
{\rm Tr}_{A} F^{A}_{\mu\nu}(x) \ast F^{A}_{\rho\sigma}(x) 
- N_{A}\, \lambda_{B}(x) \ast 
{\rm Tr}_{B} F^{B}_{\mu\nu}(x) \ast F^{B}_{\rho\sigma}(x) \nonumber \\
& & -2\lambda_{A}(x) \ast {\rm Tr}_{A} F^{A}_{\mu\nu}(x) \ast 
{\rm Tr}_{B} \int \frac{d^{4}p}{(2\pi)^{4}} 
\widetilde{F}^{B}_{\rho\sigma}(p)e^{ip \cdot x}
 \exp\left(\frac{p \circ p}{4\varepsilon} \right) \nonumber \\
& & +2\lambda_{B}(x) \ast {\rm Tr}_{A} \int \frac{d^{4}p}{(2\pi)^{4}} 
\widetilde{F}^{A}_{\mu\nu}(p)e^{ip \cdot x} 
 \exp\left(\frac{p \circ p}{4\varepsilon} \right) 
\ast {\rm Tr}_{B} F^{B}_{\rho\sigma}(x) \nonumber \\
& & +N_{A}\, \lambda_{A}(x) \ast {\rm Tr}_{B} 
\int \frac{d^{4}p}{(2\pi)^{4}} \int \frac{d^{4}q}{(2\pi)^{4}}
\widetilde{F}^{B}_{\mu\nu}(p)\widetilde{F}^{B}_{\rho\sigma}(q)
e^{ip \cdot x} \ast e^{iq \cdot x} 
\exp\left(\frac{(p+q) \circ (p+q)}{4\varepsilon} \right)  \nonumber \\
& & -N_{B}\, \lambda_{B}(x) \ast {\rm Tr}_{A} 
\int \frac{d^{4}p}{(2\pi)^{4}} \int \frac{d^{4}q}{(2\pi)^{4}}
\widetilde{F}^{A}_{\mu\nu}(p)\widetilde{F}^{A}_{\rho\sigma}(q)
e^{ip \cdot x} \ast e^{iq \cdot x}
\exp\left(\frac{(p+q) \circ (p+q)}{4\varepsilon} \right) 
\Bigl. \Bigr\} \;. \nonumber 
%
%
\end{eqnarray}
%
The factors $\exp\left(\frac{p \circ p}{4\varepsilon} \right)$ and 
$\exp\left(\frac{(p+q) \circ (p+q)}{4\varepsilon} \right)$ in the right-hand 
side of the expression (2.15) are generated from the non-planar contributions. 
The quantities $p \circ p$ and $(p+q) \circ (p+q)$ satisfy the condition 
$p \circ p \leq 0$ and $(p+q) \circ (p+q) \leq 0$ under 
the Euclidean metric, respectively. Although $p \circ p = 0$ is equivalent to 
$\widetilde{p}^{\mu}=0$, it is not equivalent to $p_{\mu}=0$ in four or 
higher dimensions. By using the notations 
$\widehat{F}^{A}_{\mu\nu}(x) \equiv 
\lim_{\varepsilon \longrightarrow 0} \int \frac{d^{4}p}{(2\pi)^{4}} 
\widetilde{F}^{A}_{\mu\nu}(p)e^{ip \cdot x} 
\exp\left(\frac{p \circ p}{4\varepsilon} \right)$ and 
$\widehat{F}^{A}_{\mu\nu}(x) \ast \widehat{F}^{A}_{\rho\sigma}(x) \equiv 
\lim_{\varepsilon \longrightarrow 0} \int \frac{d^{4}p}{(2\pi)^{4}} 
\int \frac{d^{4}q}{(2\pi)^{4}}
\widetilde{F}_{\mu\nu}(p)\widetilde{F}_{\rho\sigma}(q)
e^{ip \cdot x} \ast e^{iq \cdot x}
\exp\left(\frac{(p+q) \circ (p+q)}{4\varepsilon} \right)$, we can rewrite the 
expression (2.15) as 
\begin{eqnarray} 
& & {\cal A}^{\rm reg}_{axial}[A, B, \lambda_{A}, \lambda_{B}] = 
-\int d^{4}x \frac{1}{32\pi^{2}} 
\varepsilon^{\mu\nu\rho\sigma} \\
& & 
\Bigl\{ \Bigr. \,
N_{B}\, \lambda_{A}(x) \ast 
{\rm Tr}_{A} F^{A}_{\mu\nu}(x) \ast F^{A}_{\rho\sigma}(x) \nonumber 
- N_{A}\,\lambda_{B}(x) \ast {\rm Tr}_{B} 
F^{B}_{\mu\nu}(x) \ast F^{B}_{\rho\sigma}(x) \nonumber \\ 
& & -2\lambda_{A}(x) \ast {\rm Tr}_{A} F^{A}_{\mu\nu}(x) \cdot 
{\rm Tr}_{B} \widehat{F}^{B}_{\rho\sigma}(x) 
+2\lambda_{B}(x) \ast {\rm Tr}_{B} F^{B}_{\mu\nu}(x) \cdot 
{\rm Tr}_{A} \widehat{F}^{A}_{\rho\sigma}(x) 
\nonumber \\
& & + N_{A}\, \lambda_{A}(x) \cdot {\rm Tr}_{B} 
\widehat{F}^{B}_{\mu\nu}(x) \ast \widehat{F}^{B}_{\rho\sigma}(x)
- N_{B}\, \lambda_{B}(x) \cdot {\rm Tr}_{A} 
\widehat{F}^{A}_{\mu\nu}(x) \ast \widehat{F}^{A}_{\rho\sigma}(x)
\Bigl. \Bigr\} \;. \nonumber 
\end{eqnarray}
%
In deriving the expression (2.16), we have utilized the commutativity of the 
product: ${\rm Tr}\widehat{F}_{\mu\nu} \cdot {\rm Tr}{F}_{\rho\sigma} 
= {\rm Tr}{F}_{\rho\sigma} \cdot {\rm Tr}\widehat{F}_{\mu\nu}$. 
Note that the Moyal star product ${\rm Tr}F_{\mu\nu} \ast 
{\rm Tr}\widehat{F}_{\rho\sigma}$ (or ${\rm Tr}\widehat{F}_{\mu\nu} \ast 
{\rm Tr}{F}_{\rho\sigma}$)results in the normal (commutative) product under 
the vanishing noncommutative momentum.

The axial anomaly for the fermion in the adjoint representation can be 
obtained by setting $A_{\mu}(x)=B_{\mu}(x)$, $\lambda_{A}(x) = \lambda_{B}(x)$ 
and $N_{A}=N_{B}$ in the expression (2.16). Since the expression of anomaly is 
antisymmetric under the exchange of the subscripts A and B, the axial anomaly 
for the fermion in the adjoint representation vanishes in four dimensions.

%
\begin{flushleft}
{\bf Arbitrary even dimensions}
\end{flushleft}

Let us now return to the case of arbitrary even dimensions. The axial anomaly 
in arbitrary even dimensions are derived by taking the limit 
$\varepsilon \longrightarrow 0$, after performing the integration over 
the momentum $k_{\mu}$ in the expression (2.10). The explicit expression 
of the anomaly is given as follows,  
\begin{eqnarray} 
& & {\cal A}^{\rm reg}_{axial}[A, B, \lambda_{A}, \lambda_{B}] = 
\int d^{2n}x \frac{i^{n}}{n!(4\pi)^{n}}
\varepsilon^{\mu_{1}\mu_{2}\mu_{3}\mu_{4} \cdots 
\mu_{2n-3}\mu_{2n-2}\mu_{2n-1}\mu_{2n}} \\
& & 
\Bigl\{ \Bigr. \,
N_{B}\,\lambda_{A}(x) \ast {\rm Tr}_{A} 
F^{A}_{\mu_{1}\mu_{2}}(x) \ast F^{A}_{\mu_{3}\mu_{4}}(x) 
\ast \cdots \ast 
F^{A}_{\mu_{2n-3}\mu_{2n-2}}(x) \ast F^{A}_{\mu_{2n-1}\mu_{2n}}(x) 
\nonumber \\
& &
+ (-1)\;\;C_{\!\!\!\!\!\!\!\!\!\!n}{}_{\;\;\;\;1}\,
\lambda_{A}(x) \ast {\rm Tr}_{A} 
F^{A}_{\mu_{1}\mu_{2}}(x) \ast F^{A}_{\mu_{3}\mu_{4}}(x) 
\ast \cdots \ast 
F^{A}_{\mu_{2n-3}\mu_{2n-2}}(x) \cdot 
{\rm Tr}_{B} \widehat{F}^{B}_{\mu_{2n-1}\mu_{2n}}(x) 
\nonumber \\
& & 
+ (-1)^{2}\;\;C_{\!\!\!\!\!\!\!\!\!\!n}{}_{\;\;\;\;2}\,
\lambda_{A}(x) \ast {\rm Tr}_{A} 
F^{A}_{\mu_{1}\mu_{2}}(x) \ast F^{A}_{\mu_{3}\mu_{4}}(x) 
\ast \cdots \,\, \cdot 
{\rm Tr}_{B} \widehat{F}^{B}_{\mu_{2n-3}\mu_{2n-2}} \ast 
\widehat{F}^{B}_{\mu_{2n-1}\mu_{2n}}(x) 
\nonumber \\
& & \qquad \cdots \qquad \qquad \cdots 
\nonumber \\
& & 
+ (-1)^{n-1}\;\;C_{\!\!\!\!\!\!\!\!\!\!n}{}_{\;\;\;\;n-1}\,
\lambda_{A}(x) \ast {\rm Tr}_{A} 
F^{A}_{\mu_{1}\mu_{2}}(x) 
\cdot {\rm Tr}_{B} \widehat{F}^{B}_{\mu_{3}\mu_{4}} \ast \cdots \ast 
\widehat{F}^{B}_{\mu_{2n-3}\mu_{2n-2}} \ast 
\widehat{F}^{B}_{\mu_{2n-1}\mu_{2n}}(x) 
\nonumber \\
& & 
+ (-1)^{n}\;N_{A}\, \lambda_{A}(x) \cdot 
{\rm Tr}_{B} \widehat{F}^{B}_{\mu_{1}\mu_{2}} \ast 
\widehat{F}^{B}_{\mu_{3}\mu_{4}} \ast \cdots \ast 
\widehat{F}^{B}_{\mu_{2n-3}\mu_{2n-2}} \ast 
\widehat{F}^{B}_{\mu_{2n-1}\mu_{2n}}(x) 
\nonumber \\
& & -(-1)^{n} \times (\mbox{\rm All of the terms with subscript A 
$\leftrightarrow$ B}) \; \Bigl. \Bigr\} \;, \nonumber 
\end{eqnarray}
%
where we have used the notations $\;\;C_{\!\!\!\!\!\!\!\!\!\!n}{}_{\;\;\;\;r} 
\equiv \frac{n!}{r!(n-r)!}$ and 
\begin{eqnarray} 
& & \widehat{F}_{\mu_{1}\mu_{2}}(x) \ast \widehat{F}_{\mu_{3}\mu_{4}}(x) \ast 
\cdots \ast \widehat{F}_{\mu_{2r-1}\mu_{2r}}(x) \nonumber \\
& & \equiv \lim_{\varepsilon \longrightarrow 0} 
\int \frac{d^{2n}p}{(2\pi)^{2n}}\widetilde{F}_{\mu_{1}\mu_{2}}(p)e^{ip\cdot x} 
\ast 
\frac{d^{2n}q}{(2\pi)^{2n}}\widetilde{F}_{\mu_{1}\mu_{2}}(q)e^{iq\cdot x} 
\ast \cdots \ast 
\int \frac{d^{2n}s}{(2\pi)^{2n}}
\widetilde{F}_{\mu_{2r-1}\mu_{2r}}(s)e^{is\cdot x} \nonumber \\
& & \times 
\exp\left\{ \frac{1}{4\varepsilon} (p+q+ \cdots +s) \circ  (p+q+ \cdots +s) 
\right\}  \;.
\end{eqnarray}
%
The terms without $\widehat{F}_{\mu\nu}$ in the right-hand side of 
the expression (2.17) come from the non-planar contributions, while all the 
other terms in the same expression come from the 
non-planar contributions. 
In deriving the expression (2.17), we have utilized again the commutativity 
of the product between ${\rm Tr}(\widehat{F}_{\mu\nu} \ast 
\cdots \ast \widehat{F}_{{\rho\sigma}})$ and the other fields. 
Notice that the expression (2.17) is antisymmetric under the exchange 
of the subscript A for B in $D=4k(k=1,2,\ldots,)$ dimensions, while 
it is symmetric in $D=4k-2(k=1,2,\ldots,)$ dimensions. 
Therefore, the chiral anomalies in 
$D=4k(k=1,2,\ldots,)$ dimensions vanish in the noncommutative gauge theories 
with fermions in adjoint representation. 

In general, the limit of the cutoff parameter $\varepsilon (\sim 1/\Lambda) 
\rightarrow 0$ and that of the noncommutative momentum 
$\widetilde{p}_{\mu} \rightarrow 0$ 
do not commute in noncommutative quantum field theories \cite{SMMVR}. 
This phenomenon is known as UV/IR mixing. When we take the limit 
$\varepsilon \rightarrow 0$ after integrating over the momentum 
$k_{\mu}$ in the expression (2.10), and next take the limit 
$\widetilde{p}_{\mu} \rightarrow 0$, then we obtain the axial anomaly (2.17). 
On the other hand, when we take the limit $\widetilde{k}_{\mu} \rightarrow 0$ 
before integrating over the momentum $k_{\mu}$ in the expression (2.10), and 
next take the limit $\varepsilon \rightarrow 0$, then we obtain the axial 
anomaly comes from the planar contributions only. This phenomenon can be 
regarded as a UV/IR mixing for the axial anomalies in noncommutative gauge 
theories.

%
%
\section{The chiral gauge anomaly for bi-fundamental and adjoint 
chiral fermion}
\setcounter{equation}{0}
We next discuss the non-abelian anomalies for bi-fundamental chiral fermions 
in noncommutative Euclidean space. Let us consider a bi-fundamental chiral 
fermion $P_{R}\psi^{i}{}_{j}(x) \equiv \frac{1+\gamma_{2n+1}}{2}
\psi^{i}{}_{j}(x)$ interacting with a $U(N_{A})$ gauge field 
$A_{\mu}{}^{i_{1}}{}_{i_{2}}(x)$ and a $U(N_{B})$ gauge field 
$B_{\mu}{}^{j_{2}}{}_{j_{1}}(x)$. The classical action on 2n-dimensional 
noncommutative (Euclidean) space is given by 
\begin{eqnarray}
S[\bar{\psi}, \psi, A, B] = \int d^{2n}x \; \bar{\psi}^{j}{}_{i}(x) \ast 
(i /\!\!\!\!D_{R}[A, B]) \psi^{i}{}_{j} (x) \;, 
\label{eqn:(3.1)}
\end{eqnarray}
%
with the Dirac operator $/\!\!\!\!D_{R}[A, B]$  
\begin{eqnarray}
& & i/\!\!\!\!D_{R}[A, B]\psi^{i_{1}}{}_{j_{1}}(x) 
= (\; /\!\!\!\partial \delta^{i_{1}}{}_{i_{2}}\delta^{j_{2}}{}_{j_{1}} 
+ A_{\mu}{}^{i_{1}}{}_{i_{2}}(x) \ast \delta^{j_{2}}{}_{j_{1}} 
\gamma^{\mu}P_{R}
- \delta^{i_{1}}{}_{i_{2}} \ast B_{\mu}{}^{j_{2}}{}_{j_{1}}(x) 
\gamma^{\mu}P_{R}
\;) \psi^{i_{2}}{}_{j_{2}}(x) \;. \nonumber \\
& & 
\label{eqn:(3.2)}
\end{eqnarray}
%
Here the index $i$ runs from 1 to $N_{A}$ and $j$ from 1 to $N_{B}$, 
respectively. Although the Dirac operator $/\!\!\!\!D_{R}[A, B]$ is not 
Hermitian in Euclidean space, the operators 
$/\!\!\!\!D_{R}^{\dagger}/\!\!\!\!D_{R}$ and 
$/\!\!\!\!D_{R}/\!\!\!\!D_{R}^{\dagger}$ are Hermitian and 
positive definite: 
\begin{eqnarray} 
/\!\!\!\!D_{R}^{\dagger}/\!\!\!\!D_{R}\varphi_{n}(x) 
= \lambda_{n}^{2} \varphi_{n}(x) \; ,  \qquad 
/\!\!\!\!D_{R}/\!\!\!\!D_{R}^{\dagger}\phi_{n}(x) 
= \lambda_{n}^{2} \phi_{n}(x)\;, \nonumber 
\end{eqnarray}
then we can introduce the orthonormal and complete systems 
$\{\,\varphi_{n}(x)\,\}$ and $\{\,\phi_{n}(x)\,\}$:
\begin{eqnarray} 
& & \int d^{2n}x \, \varphi^{\dagger}_{m}(x) \, \varphi_{n}(x) 
= \int d^{2n}x \, \phi^{\dagger}_{m}(x) \, \phi_{n}(x) = \delta_{mn} \;. 
\end{eqnarray}
%
The infinitesimal gauge transformations for the fermions are given as 
follows, 
\begin{eqnarray}
& & \delta \psi^{i_{1}}{}_{j_{1}}(x) = 
\Lambda_{A}{}^{i_{1}}{}_{i_{2}}(x) \ast P_{R} \psi^{i_{2}}{}_{j_{1}}(x) 
 - P_{R} \psi^{i_{1}}{}_{j_{2}}(x) \ast \Lambda_{B}{}^{j_{2}}{}_{j_{1}}(x) 
\;,  \nonumber \\ 
& &  \\ 
& & \delta \bar{\psi}^{j_{1}}{}_{i_{1}}(x) = 
- \bar{\psi}^{j_{1}}{}_{i_{2}}(x)P_{L} \ast \Lambda_{A}{}^{i_{2}}{}_{i_{1}}(x) 
+ \Lambda_{B}{}^{j_{1}}{}_{j_{2}}(x) \ast \bar{\psi}^{j_{2}}{}_{i_{1}}(x)P_{L} 
\;, \nonumber 
\end{eqnarray}
%
with $P_{L} \equiv \frac{1-\gamma_{2n+1}}{2}$. 
Here $\Lambda_{A}{}^{i_{1}}{}_{i_{2}}(x)$ and 
$\Lambda_{B}{}^{i_{1}}{}_{i_{2}}(x)$ denote some infinitesimal functions. 
Let us consider the effective action for the gauge fields derived from the 
classical action (3.1): 
\begin{eqnarray}
W[A, B] = \ln \int {\cal D}\bar{\psi} {\cal D}\psi 
\exp \left( -\int d^{2n}x \; \bar{\psi}^{j}{}_{i}(x) \ast 
(i /\!\!\!\!D_{R}[A, B]) \psi^{i}{}_{j} (x) \right) \;,  
\label{eqn:(3.5)}
\end{eqnarray}
%
The invariance of the effective action (3.5) under the infinitesimal 
transformations (3.4) leads 
\begin{eqnarray}
& & \int d^{2n}x \left\{ \Lambda_{A}{}^{i_{1}}{}_{i_{2}}(x) \ast 
(D^{(A)}_{R}{}_{\mu}J^{\mu (A)}_{2n+1}){}^{i_{2}}{}_{i_{1}}(x) 
+ \Lambda_{B}{}^{j_{1}}{}_{j_{2}}(x) \ast 
(D^{(B)}_{R}{}_{\mu}J^{\mu (B)}_{2n+1}){}^{j_{2}}{}_{j_{1}}(x) \right\} 
\nonumber \\ 
& & = {\cal A}_{chiral}[A, B, \Lambda_{A}, \Lambda_{B}] \;, 
\end{eqnarray}
%
where $J^{\mu (A)}_{2n+1}{}^{i_{1}}{}_{i_{2}}(x)$ and 
$J^{\mu (B)}_{2n+1}{}^{j_{1}}{}_{j_{2}}(x)$ are the (right-handed) nonabelian 
currents: 
\begin{eqnarray}
& & J^{\mu (A)}_{2n+1}{}^{i_{1}}{}_{i_{2}}(x) \equiv 
(\psi_{\beta}{}^{i_{1}}{}_{j_{1}}(x) \ast 
\bar{\psi}_{\alpha}{}^{j_{1}}{}_{i_{2}}(x))
(\gamma^{\mu}P_{R})_{\alpha\beta} \;,  \nonumber \\ 
& & \\
& & J^{\mu (B)}_{2n+1}{}^{j_{1}}{}_{j_{2}}(x) \equiv 
(\bar{\psi}_{\alpha}{}^{j_{1}}{}_{i_{1}}(x) \ast 
\psi_{\beta}{}^{i_{1}}{}_{j_{2}}(x))
(\gamma^{\mu}P_{R})_{\alpha\beta} \;.  \nonumber 
\end{eqnarray}
%
and the covariant derivative of these current are given by 
\begin{eqnarray}
& & D^{(A)}_{\mu}J^{\mu (A)}_{2n+1}{}^{i_{1}}{}_{i_{2}}(x) \equiv 
\partial_{\mu} J^{\mu (A)}_{2n+1}{}^{i_{1}}{}_{i_{2}}(x) 
+ A_{\mu}{}^{i_{1}}{}_{i_{3}}(x) 
\ast J^{\mu (A)}_{2n+1}{}^{i_{3}}{}_{i_{2}}(x) 
- J^{\mu (A)}_{2n+1}{}^{i_{1}}{}_{i_{3}}(x) \ast 
A_{\mu}{}^{i_{3}}{}_{i_{2}}(x) \;, \nonumber \\ 
& & \nonumber \\
& & D^{(B)}_{\mu}J^{\mu (B)}_{2n+1}{}^{j_{1}}{}_{j_{2}}(x) \equiv 
\partial_{\mu} J^{\mu (B)}_{2n+1}{}^{j_{1}}{}_{j_{2}}(x) 
+ B_{\mu}{}^{j_{1}}{}_{j_{3}}(x) 
\ast J^{\mu (B)}_{2n+1}{}^{j_{3}}{}_{j_{2}}(x) 
- J^{\mu (B)}_{2n+1}{}^{j_{1}}{}_{j_{3}}(x) \ast 
B_{\mu}{}^{j_{3}}{}_{j_{2}}(x) \;. \nonumber 
\end{eqnarray}
The right-hand side of the expression (3.6) is contained in the Jacobian 
factor of the path integral measure in the effective action (3.5). 
We evaluate the Jacobian factor with respect to the gauge transformation of 
the fermions. Under the infinitesimal gauge transformation 
the path integral measure in the effective action (3.5) transforms as 
${\cal D}\widetilde{\psi} {\cal D}\widetilde{\bar{\psi}} = 
J_{chiral}[\Lambda_{A}, \Lambda_{B}]{\cal D}\psi {\cal D}\bar{\psi}$ with the 
Jacobian 
\begin{eqnarray} 
J_{chiral}[\Lambda_{A}, \Lambda_{B}] 
= \exp\left( - {\cal A}_{chiral}[\Lambda_{A}, \Lambda_{B}] \right) \;. 
\end{eqnarray}
%
Since the Jacobian is ill defined, we have to regularize it by inserting a 
damping factor. Then we obtain two types of chiral anomaly reflecting 
the two different regularization procedures. We perform the regularization 
in a gauge covariant way \cite{KF, CPM2}. 
Inserting the Gaussian factor 
$\exp(-\varepsilon /\!\!\!\!D_{R}^{\dagger}/\!\!\!\!D_{R})$ and 
$\exp(-\varepsilon /\!\!\!\!D_{R}/\!\!\!\!D_{R}^{\dagger})$, we have 
\begin{eqnarray} 
& & {\cal A}^{\rm reg}_{chiral}[A, B, \Lambda_{A}, \Lambda_{B}] \\
\!\!&=&\!\! \lim_{\varepsilon \longrightarrow 0} \Sigma_{n} \int d^{2n}x 
\Bigl[ \Bigr. \, 
\Lambda_{A}(x) \ast \left( P_{R} 
( e^{-\varepsilon /\!\!\!\!D_{R}^{\dagger}/\!\!\!\!D_{R}} 
\varphi_{n}(x) ) \ast \varphi^{\dagger}_{n}(x) 
- P_{L} 
( e^{-\varepsilon /\!\!\!\!D_{R}/\!\!\!\!D_{R}^{\dagger}} 
\phi_{n}(x) ) \ast \phi^{\dagger}_{n}(x) \right) \nonumber \\
\!\!&-&\!\! \Lambda_{B}(x) \ast \left( 
\varphi^{\dagger}_{n}(x) \ast P_{R}
( e^{-\varepsilon /\!\!\!\!D_{R}^{\dagger}/\!\!\!\!D_{R}} 
\varphi_{n}(x) ) 
- \phi^{\dagger}_{n}(x) \ast P_{L}
( e^{-\varepsilon /\!\!\!\!D_{R}/\!\!\!\!D_{R}^{\dagger}} 
\phi_{n}(x) ) \right) \, 
\Bigl. \Bigr] \;. \nonumber \\
\!\!&=&\!\! \lim_{\varepsilon \longrightarrow 0} \Sigma_{n} \int d^{2n}x 
\Bigl[ \Bigr. \, 
\Lambda_{A}(x) \ast \left( P_{R} 
( e^{\varepsilon /\!\!\!\!D^{2}} 
\varphi_{n}(x) ) \ast \varphi^{\dagger}_{n}(x) 
- P_{L} 
( e^{\varepsilon /\!\!\!\!D^{2}} 
\phi_{n}(x) ) \ast \phi^{\dagger}_{n}(x) \right) \nonumber \\
\!\!&-&\!\! \Lambda_{B}(x) \ast \left( 
\varphi^{\dagger}_{n}(x) \ast P_{R}
( e^{\varepsilon /\!\!\!\!D^{2}} 
\varphi_{n}(x) )
- \phi^{\dagger}_{n}(x) \ast P_{L}
( e^{\varepsilon /\!\!\!\!D^{2}} 
\phi_{n}(x) ) \right) \, 
\Bigl. \Bigr] \; , \nonumber 
\end{eqnarray}
%
where the differential operator $/\!\!\!\!D$ is the Dirac operator given in 
the expression (2.2). In deriving the expression (3.9), we have used the 
property of the projection operators $P_{R}$ and $P_{L}$. 

We now evaluate the regularized sum in Fourier space 
$\displaystyle{\varphi_{n}(x)=\int\frac{d^{2n}k}{(2\pi)^{2n}}e^{ik\cdot x} 
\widetilde{\varphi}_{n}(k)}$. Then we have 
\begin{eqnarray} 
& & {\cal A}^{\rm reg}_{chiral}[A, B, \Lambda_{A}, \Lambda_{B}] = 
\lim_{\varepsilon \longrightarrow 0} 
\int d^{2n}x \int \frac{d^{2n}k}{(2\pi)^{n}}  \\
& & \mbox{\boldmath ${\rm Tr}$} 
\Bigl[ \Bigr. \,  \Lambda_{A}(x) \ast P_{R} 
\exp\left( -\varepsilon ( i /\!\!\!k +/\!\!\!\!D )^{2} \right) 
e^{ik \cdot x} \ast e^{-ik \cdot x} 
- \Lambda_{A}(x) \ast P_{L} 
\exp\left( -\varepsilon ( i /\!\!\!k +/\!\!\!\!D )^{2} \right) 
e^{ik \cdot x} \ast e^{-ik \cdot x}  \nonumber \\
& & - \Lambda_{B}(x) \ast  
e^{-ik \cdot x} \ast P_{R}
\exp\left( -\varepsilon ( i /\!\!\!k +/\!\!\!\!D )^{2} \right)
e^{ik \cdot x} 
+ \Lambda_{B}(x) \ast e^{-ik \cdot x} \ast P_{L}
e^{\varepsilon /\!\!\!\!D^{2}} 
\exp\left( -\varepsilon ( i /\!\!\!k +/\!\!\!\!D )^{2} \right) 
e^{ik \cdot x} \, \Bigl. \Bigr] \; , \nonumber 
\end{eqnarray}
%
where the notation ``$\mbox{\boldmath ${\rm Tr}$}$" denotes the trace over the 
$U(N_{A})$, $U(N_{B})$ and the Dirac matrices. Inserting the 
decomposition $/\!\!\!\!D^{2} = D_{\mu}D^{\mu} 
+ \frac{1}{2}\gamma^{\mu}\gamma^{\nu}
[\,D_{\mu}, \, D_{\nu} \,]$ into the expression (3.10) and performing the 
rescaling $k_{\mu} \longrightarrow (1/\sqrt{\varepsilon} k_{\mu})$, we have  
\begin{eqnarray} 
& & {\cal A}^{\rm reg}_{chiral}[A, B, \Lambda_{A}, \Lambda_{B}] = 
\lim_{\varepsilon \longrightarrow 0} 
\int d^{2n}x \int \frac{d^{2n}k}{(2\pi)^{n}} e^{k \cdot k}
\frac{i^{n}}{n!} \varepsilon^{\mu_{1}\mu_{2} \cdots \mu_{2n}} \\
& &  \Bigl\{ \Bigr.
{\rm Tr}_{A}{\rm Tr}_{B} \, \left[ \, \Lambda_{A}(x) \ast 
(F^{A}_{\mu_{1}\mu_{2}}(x) \ast 
- \ast F^{B}_{\mu_{1}\mu_{2}}(x+\frac{\widetilde{k}}{\sqrt{\varepsilon}}))
 \cdots (F^{A}_{\mu_{2n-1}\mu_{2n}}(x) \ast 
-  \ast F^{B}_{\mu_{2n-1}\mu_{2n}}(x+\frac{\widetilde{k}}{\sqrt{\varepsilon}})) \right] \nonumber \\ 
& & - {\rm Tr}_{A}{\rm Tr}_{B} \, \left[ \, \Lambda_{B}(x) \ast 
(F^{A}_{\mu_{1}\mu_{2}}(x-\frac{\widetilde{k}}{\sqrt{\varepsilon}}) \ast 
- \ast F^{B}_{\mu_{1}\mu_{2}}(x))
\cdots (F^{A}_{\mu_{2n-1}\mu_{2n}}(x-\frac{\widetilde{k}}{\sqrt{\varepsilon}}) \ast 
-  \ast F^{B}_{\mu_{2n-1}\mu_{2n}}(x)) 
\right]   \Bigl. \Bigr\} \;, \nonumber 
\end{eqnarray}
%
where the notations ${\rm Tr}_{A}$ and ${\rm Tr}_{B}$ denote the traces over 
the $U(N_{A})$ and $U(N_{B})$ matrices, respectively. 
We shall advance to the calculation in arbitrary dimensions after examining 
the case of two and four dimensions. 

%
%
\begin{flushleft}
{\bf  Two dimensions}
\end{flushleft}
The explicit form of the expression (3.11) in two dimensions is given by 
\begin{eqnarray} 
& & {\cal A}^{\rm reg}_{chiral}[A, B, \Lambda_{A}, \Lambda_{B}] = 
\lim_{\varepsilon \longrightarrow 0} 
\int d^{2}x \int \frac{d^{2}k}{(2\pi)^{n}} e^{k \cdot k}
\frac{-1}{2} \varepsilon^{\mu\nu} \\ 
& & \Bigl\{ \Bigr.
N_{B}\,{\rm Tr}_{A} \Lambda_{A}(x) \ast F^{A}_{\mu\nu}(x) 
+ N_{A}\,{\rm Tr}_{B} \Lambda_{B}(x) \ast F^{B}_{\mu\nu}(x) \nonumber \\
& & - {\rm Tr}_{A} \Lambda_{A}(x) \ast 
{\rm Tr}_{B} F^{B}_{\mu\nu}(x-\frac{\widetilde{k}}{\sqrt{\varepsilon}}) 
- {\rm Tr}_{B} \Lambda_{B}(x) \ast 
{\rm Tr}_{A} F^{A}_{\mu\nu}(x-\frac{\widetilde{k}}{\sqrt{\varepsilon}}) 
\Bigl. \Bigr\} \;, \nonumber 
\end{eqnarray}
%
where the coefficients $N_{A}$ and $N_{B}$ arise from $N_{A} \equiv 
{\rm Tr}_{A} I_{N_{A} \times N_{A}}$ and $N_{B} \equiv 
{\rm Tr}_{B} I_{N_{B} \times N_{B}}$, respectively. Performing the 
integration over the momentum $k$, we find that the factor 
$\exp(\frac{1}{4\varepsilon}p \circ p)$ is generated from the non-planar 
contributions. Since $p \circ p$ takes a negative value for arbitrary non-zero 
momentum $p_{\mu}$ in two dimensions, we obtain the following result under the 
limit $\varepsilon \longrightarrow 0$, 
\begin{eqnarray} 
{\cal A}^{\rm reg}_{chiral}[A, B, \Lambda_{A}, \Lambda_{B}] \!\!&=&\!\! 
\int d^{2}x \frac{1}{4\pi} 
i \varepsilon^{\mu\nu} 
\Bigl\{ \Bigr. \,
N_{B}\, {\rm Tr}_{A} \Lambda_{A}(x) \ast F^{A}_{\mu\nu}(x) \ast 
+ N_{A}\,{\rm Tr}_{B} \Lambda_{B}(x) \ast F^{B}_{\mu\nu}(x) \nonumber \\ 
\!\!&-&\!\! 
{\rm Tr}_{A}\Lambda_{A}(x) \, {\rm Tr}_{B} \widetilde{F}^{B}_{\mu\nu}(0) 
- {\rm Tr}_{B}\Lambda_{B}(x) \, {\rm Tr}_{A} \widetilde{F}^{A}_{\mu\nu}(0) 
\Bigl. \Bigr\} \;.
\end{eqnarray}
%
If the Fourier coefficients $\widetilde{F}^{A}_{\mu\nu}(0)$ and 
$\widetilde{F}^{B}_{\mu\nu}(0)$ are non-zero, there are non-planar 
contributions to the chiral gauge anomaly. 

%
\begin{flushleft}
{\bf Four dimensions}
\end{flushleft}
The chiral gauge anomaly in four dimensions is also derived by taking the 
limit $\varepsilon \longrightarrow 0$ after performing the integration over 
the momentum $k_{\mu}$ in the expression (3.11) with $n=2$. The concrete form 
of the chiral gauge anomaly is given by 
\begin{eqnarray} 
& & {\cal A}^{\rm reg}_{chiral}[A, B, \Lambda_{A}, \Lambda_{B}] = 
-\int d^{4}x \frac{1}{(2\pi)^{4}} 
\varepsilon^{\mu\nu\rho\sigma} \\
& & 
\Bigl\{ \Bigr. \,
N_{B}\,{\rm Tr}_{A} \Lambda_{A}(x) \ast 
F^{A}_{\mu\nu}(x) \ast F^{A}_{\rho\sigma}(x)  
- N_{A}\,{\rm Tr}_{B} \Lambda_{B}(x) \ast 
F^{B}_{\mu\nu}(x) \ast F^{B}_{\rho\sigma}(x)  
\nonumber \\ 
& & - 2 \,{\rm Tr}_{A} \Lambda_{A}(x) \ast F^{A}_{\mu\nu}(x) 
\cdot {\rm Tr}_{B} \widehat{F}^{B}_{\rho\sigma}(x) 
+2 \,{\rm Tr}_{B} \Lambda_{B}(x) \ast F^{B}_{\mu\nu}(x) 
\cdot {\rm Tr}_{A} \widehat{F}^{A}_{\rho\sigma}(x) \nonumber \\
& & + {\rm Tr}_{A} \Lambda_{A}(x) \cdot 
{\rm Tr}_{B} \widehat{F}^{B}_{\mu\nu}(x) \ast \widehat{F}^{B}_{\rho\sigma}(x)
- {\rm Tr}_{B} \Lambda_{B}(x) \cdot 
{\rm Tr}_{A} \widehat{F}^{A}_{\mu\nu}(x) \ast \widehat{F}^{A}_{\rho\sigma}(x)
\Bigl. \Bigr\} \;, \nonumber 
\end{eqnarray}
%
where we have used the notations introduced in the expression (2.18). 
In deriving the expression (3.14), we have utilized the commutativity of 
the product between ${\rm Tr}\widehat{F}_{\mu\nu}$ and the other fields. 
We see that the mixed $U(N)U(M)^{2}$ or $U(N)^{2}U(M)$ anomaly comes from the 
non-planar contributions. 
The chiral gauge anomaly for an adjoint fermion can be obtained by setting 
$A_{\mu}(x)=B_{\mu}(x)$, $\Lambda_{A}(x) = \Lambda_{B}(x)$ and $N_{A}=N_{B}$ 
in the expression (3.14). 
Since the expression of anomaly is antisymmetric under the exchange of 
subscript A and B, the chiral gauge anomaly vanish in four dimensions.  
Therefore noncommutative gauge theories with adjoint chiral 
fermions are anomaly free in four dimensions. 

%
\begin{flushleft}
{\bf Arbitrary even dimensions}
\end{flushleft}
Let us now return to the arbitrary even dimensions. By the same calculations 
as the case of two and four dimensions, we obtain the concrete form of the 
chiral gauge anomaly in arbitrary even dimensions:

\begin{eqnarray} 
& & {\cal A}^{\rm reg}_{axial}[A, B, \Lambda_{A}, \Lambda_{B}] = 
\int d^{2n}x \frac{i^{n}}{n!(4\pi)^{n}} 
\varepsilon^{\mu_{1}\mu_{2}\mu_{3}\mu_{4} \cdots 
\mu_{2n-3}\mu_{2n-2}\mu_{2n-1}\mu_{2n}} \\
& & 
\Bigl\{ \Bigr. \,
N_{B}\,{\rm Tr}_{A} \Lambda_{A}(x) \ast 
F^{A}_{\mu_{1}\mu_{2}}(x) \ast F^{A}_{\mu_{3}\mu_{4}}(x) 
\ast \cdots \ast 
F^{A}_{\mu_{2n-3}\mu_{2n-2}}(x) \ast F^{A}_{\mu_{2n-1}\mu_{2n}}(x) 
\nonumber \\
& &
+ (-1)\;\;C_{\!\!\!\!\!\!\!\!\!\!n}{}_{\;\;\;\;1}\,
{\rm Tr}_{A} \Lambda_{A}(x) \ast 
F^{A}_{\mu_{1}\mu_{2}}(x) \ast F^{A}_{\mu_{3}\mu_{4}}(x) 
\ast \cdots \ast 
F^{A}_{\mu_{2n-3}\mu_{2n-2}}(x) \cdot 
{\rm Tr}_{B} \widehat{F}^{B}_{\mu_{2n-1}\mu_{2n}}(x) 
\nonumber \\
& & 
+ (-1)^{2}\;\;C_{\!\!\!\!\!\!\!\!\!\!n}{}_{\;\;\;\;2}\,
{\rm Tr}_{A} \Lambda_{A}(x) \ast 
F^{A}_{\mu_{1}\mu_{2}}(x) \ast F^{A}_{\mu_{3}\mu_{4}}(x) 
\ast \cdots \,\, \cdot 
{\rm Tr}_{B} \widehat{F}^{B}_{\mu_{2n-3}\mu_{2n-2}} \ast 
\widehat{F}^{B}_{\mu_{2n-1}\mu_{2n}}(x) 
\nonumber \\
& & \qquad \cdots \qquad \qquad \cdots 
\nonumber \\
& & 
+ (-1)^{n-1}\;\;C_{\!\!\!\!\!\!\!\!\!\!n}{}_{\;\;\;\;n-1}\,
{\rm Tr}_{A} \Lambda_{A}(x) \ast 
F^{A}_{\mu_{1}\mu_{2}}(x) 
\cdot {\rm Tr}_{B} \widehat{F}^{B}_{\mu_{3}\mu_{4}} \ast \cdots \ast 
\widehat{F}^{B}_{\mu_{2n-3}\mu_{2n-2}} \ast 
\widehat{F}^{B}_{\mu_{2n-1}\mu_{2n}}(x) 
\nonumber \\
& & 
+ (-1)^{n}\;\;C_{\!\!\!\!\!\!\!\!\!\!n}{}_{\;\;\;\;n}\,
{\rm Tr}_{A} \Lambda_{A}(x) \cdot 
{\rm Tr}_{B} \widehat{F}^{B}_{\mu_{1}\mu_{2}} \ast 
\widehat{F}^{B}_{\mu_{3}\mu_{4}} \ast \cdots \ast 
\widehat{F}^{B}_{\mu_{2n-3}\mu_{2n-2}} \ast 
\widehat{F}^{B}_{\mu_{2n-1}\mu_{2n}}(x) 
\nonumber \\
& & -(-1)^{n} \times (\mbox{\rm All of the terms with subscript A 
$\leftrightarrow$ B}) \; \Bigl. \Bigr\} \;, \nonumber 
\end{eqnarray}
%
with the notation given in the expression (2.18). 
Here we have utilized again the commutativity 
of the product between ${\rm Tr}(\widehat{F}_{\mu\nu} \ast 
\cdots \ast \widehat{F}_{{\rho\sigma}})$ and the other fields. 
The terms without $\widehat{F}_{\mu\nu}$ in the right-hand side 
of the expression (3.15) come from the non-planar contributions, while all the 
other terms in the same expression come from the non-planar contributions. 

We notice the expression (3.15) is antisymmetric under the exchange of 
the subscript A for B in $D=4k(k=1,2,\ldots,)$ dimensions, while it is 
symmetric in $D=4k-2(k=1,2,\ldots,)$ dimensions. Therefore, the chiral 
anomalies in $D=4k(k=1,2,\ldots,)$ dimensions vanish in the theories with 
adjoint chiral fermions. Namely, the noncommutative gauge theories with 
adjoint chiral fermions are anomaly free in $D=4k(k=1,2,\ldots,)$ dimensions.

When we take the limit 
$\varepsilon \rightarrow 0$ after integrating over the momentum 
$k_{\mu}$ in the expression (3.11), and next take the limit 
$\widetilde{p}_{\mu} \rightarrow 0$, then we obtain the chiral gauge anomaly 
(3.15). 
On the other hand, when we take the limit $\widetilde{k}_{\mu} \rightarrow 0$ 
before integrating over the momentum $k_{\mu}$ in the expression (3.11), and 
next take the limit $\varepsilon \rightarrow 0$, then we obtain the chiral 
gauge anomaly comes from the planar contributions only. This phenomenon can be 
regarded as a UV/IR mixing for the chiral gauge anomalies in 
noncommutative gauge theories.

%
\section{Conclusions}
\setcounter{equation}{0}

In this paper, we have calculated the axial and chiral gauge anomalies 
emerging from non-planar sector in noncommutative gauge theories. 
In noncommutative gauge theories with fermions in the bi-fundamental and 
the adjoint representation, there are non-planar contributions to the 
anomalies. These anomalies in non-planar sector can be evaluated not only 
in perturbative analysis but also in path integral formulation. 
When the regularization by introducing a Gaussian 
cut-off is performed in path integral formulation, non-planar sector includes 
the damping factor depending on the noncommutative momentum 
$\widetilde{p}^{\mu} \equiv p_{\nu}\theta^{\mu\nu}$. 
Therefore, the anomaly with the non-zero noncommutative momentum in non-planar 
sector vanishes. This fact has been shown also by the perturbative analysis 
about the chiral gauge anomalies \cite{CPM2}. The argument, however, breaks 
down for the zero noncommutative momentum, since in this case the non-planar 
sector is not regularized by the damping factor. Therefore the anomalies in 
non-planar sector remain for the zero noncommutative momentum. This result is 
consistent with the result obtained in perturbative analysis \cite{AAELST}.

In the noncommutative gauge theories with adjoint chiral fermion, 
the chiral gauge anomalies in planar sector vanish in $D=4k(k=1,2,\ldots,)$ 
dimensions \cite{LBMSAT, CPM2}. 
The adjoint chiral fermion can be regarded as the product of fundamental and 
anti-fundamental chiral fermions. Since the fundamental and 
anti-fundamental chiral fermions give opposite contributions to 
the chiral gauge anomalies in $D=4k(k=1,2,\ldots,)$ dimensions, the 
anomalies cancel out. The cancellation mechanism is also valid in non-planar 
sector. The chiral gauge anomalies in non-planar sector cancel out 
in $D=4k(k=1,2,\ldots,)$ dimensions.

Noncommutative quantum field theories exhibit an intriguing mixing of the 
ultraviolet and infrared regions. 
The limit of the cut-off parameter $\varepsilon \rightarrow 0$ and that of 
the noncommutative momentum $\widetilde{p}^{\mu} \rightarrow 0$ 
do not commute. In deriving the axial and chiral gauge 
anomaly by path integral formulation, we find that the limits of 
the cutoff parameter and that of the noncommutative momentum do not 
commute, either. This phenomenon can be interpreted as UV/IR mixing 
for the anomalies in noncommutative gauge theories. Namely, 
the IR singularity (at the vanishing noncommutative momentum) in 
non-planar sector leads to the anomalies on the UV behavior via the 
intriguing UV/IR mixing in noncommutative gauge theories.

It was well known that there are two types of chiral gauge anomaly: the 
covariant anomaly and the consistent anomaly. Although the consistent anomaly 
is not gauge covariant, it is a solution of the Wess--Zumino consistency 
condition. Hence the cohomological method is applicable to deriving the 
consistent anomaly \cite{LBMSAT, LBAS}. The consistent anomaly and the 
covariant anomaly are related to the Moyal star polynomial of the 
Bardeen--Zumino type \cite{CPM1, CPM2}. It will be an interesting 
subject to investigate the Moyal star polynomial of the Bardeen--Zumino type 
coming from non-planar contributions in noncommutative gauge theories. We 
hope to discuss this subject in the future.


\section*{Acknowledgments}

I am grateful to S. Deguchi for helpful suggestions. 
I would also like to thank A. Sugamoto for careful reading of the manuscript 
and useful comments.


\clearpage


\end{document}